%% file: root.tex
\PassOptionsToPackage{doi=false,isbn=false,url=false}{bibtex}
\documentclass[letterpaper, 10 pt, conference]{ieeeconf} 

\IEEEoverridecommandlockouts                              

\usepackage{graphicx} 
\graphicspath{ {./figures/} }
\usepackage{amsmath} 
\usepackage{amssymb}
\usepackage{algorithm}
\usepackage[noend]{algpseudocode}
\usepackage{xcolor}
\definecolor{tumblue}{RGB}{0,101,189}

\usepackage{stfloats}
\usepackage{setspace}
\usepackage{tabularx}
\usepackage{booktabs}
\usepackage[T1]{fontenc}
\usepackage[scaled=0.85]{beramono}
\usepackage{listings}
\usepackage{caption}
\usepackage{cleveref}
\Crefname{lstlisting}{Listing}{Listings}
\usepackage{listings}
\lstdefinestyle{sparqlstyle}{
  language=OCL,
  basicstyle=\footnotesize,
  stepnumber=1,
  numbersep=10pt,
  tabsize=2,
  showspaces=false,
  breaklines=true
}

\usepackage[nonumberlist,nopostdot,acronym,toc]{glossaries}
\newacronym{oee}{OEE}{Overall Equipment Effectiveness}
\newacronym{abox}{ABox}{assertional component}
\newacronym{aps}{aPS}{automated Production System}
\newacronym{cpsos}{CPSoS}{Cyber-Physical Systems of Systems}
\newacronym{cps}{CPS}{Cyber-Physical System}
\newacronym{kb}{KB}{Knowledge Base}
\newacronym{mas}{MAS}{Multi Agent System}
\newacronym{owl}{OWL}{Web Ontology Language}
\newacronym{pa}{PA}{product agent}
\newacronym{oa}{OA}{ontology agent}
\newacronym{ppr}{PPR}{Product Process Resource}
\newacronym{rdf}{RDF}{Resource Description Framework}
\newacronym{ra}{RA}{resource agent}
\newacronym{sparql}{SPARQL}{SPARQL Protocol And RDF Query Language}
\newacronym{swt}{SWT}{Semantic Web Technologies}
\newacronym{tbox}{TBox}{terminological component}
\newacronym{api}{API}{Application Programming Interface}
\newacronym{rest}{REST}{REpresentational State Transfer}
\newacronym{json}{JSON}{JavaScript Object Notation}
\newacronym{http}{HTTP}{HyperText Transfer Protocol}
\newacronym{bdi}{BDI}{Belief-Desire-Intention}
\newacronym{adacor}{ADACOR}{ADAptive holonic COntrol aRchitecture}

\usepackage{balance} 

\makeatletter
\newcommand\fs@betterruled{%
  \def\@fs@cfont{\bfseries}\let\@fs@capt\floatc@ruled
  \def\@fs@pre{\vspace*{5pt}\hrule height.8pt depth0pt \kern2pt}%
  \def\@fs@post{\kern2pt\hrule\relax}%
  \def\@fs@mid{\kern2pt\hrule\kern2pt}%
  \let\@fs@iftopcapt\iftrue}
\floatstyle{betterruled}
\restylefloat{algorithm}
\makeatother

\author{Jonghan Lim$^{1}$, Leander Pfeiffer$^{2}$, Felix Ocker$^{2}$, Birgit Vogel-Heuser$^{2}$, and Ilya Kovalenko$^{1}$
\thanks{$^{1}$Jonghan Lim and Ilya Kovalenko are with Department of Industrial and Manufacturing and Department of Mechanical Engineering,
        Pennsylvania State University, State College, USA
        (e-mail: \{jxl567; iqk5135\}@psu.edu).
        }%
\thanks{$^{2}$Leander Pfeiffer, Felix Ocker and Birgit Vogel-Heuser are with Institute of Automation and Information Systems,
        Technical University of Munich, Munich, Germany
        (e-mail: \{leander.pfeiffer; felix.ocker; vogel-heuser\}@tum.de).
        }%
}

\title{\LARGE \bf
Ontology-Based Feedback to Improve Runtime Control for Multi-Agent Manufacturing Systems
}

\begin{document}

\maketitle
\thispagestyle{empty}
\pagestyle{empty}



\begin{abstract}
Improving the overall equipment effectiveness (OEE) of machines on the shop floor is crucial to ensure the productivity and efficiency of manufacturing systems. To achieve the goal of increased OEE, there is a need to develop flexible runtime control strategies for the system. Decentralized strategies, such as multi-agent systems, have proven effective in improving system flexibility. However, runtime multi-agent control of complex manufacturing systems can be challenging as the agents require extensive communication and computational efforts to coordinate agent activities. One way to improve communication speed and cooperation capabilities between system agents is by providing a common language between these agents to represent knowledge about system behavior. The integration of ontology into multi-agent systems in manufacturing provides agents with the capability to continuously update and refine their knowledge in a global context. This paper contributes to the design of an ontology for multi-agent systems in manufacturing, introducing an extendable knowledge base and a methodology for continuously updating the production data by agents during runtime. To demonstrate the effectiveness of the proposed framework, a case study is conducted in a simulated environment, which shows improvements in OEE during runtime.
\end{abstract}

\section{Introduction}
\label{sec:introduction}

\input{10_introduction.tex}

\section{Background}
\label{sec:background}

\input{20_stateoftheart.tex}

\section{Framework for Runtime Control of Agents}
\label{sec:framework}

\input{30_framework.tex}

\section{Implementation}
\label{sec:implementation}

\input{40_implementation.tex}

\section{Case Study}
\label{sec:casestudy}

\input{50_casestudy.tex}

\section{Summary and Outlook}
\label{sec:summary}

\input{60_summary.tex}

\balance

\bibliographystyle{IEEEtran}
%

\bibliography{bib_PA_SWT_3}






\end{document}

%% file: 10_introduction.tex

Manufacturing systems have become increasingly complex in recent years due to various factors, including globalization, the spread of new technology, the rising demand for customized products, and the growing concern about sustainability. As a result, manufacturers are facing the challenge of managing the performance of their manufacturing systems to ensure that they operate efficiently and effectively, which can be achieved by improving \gls{oee}~\cite{nakajima1988introduction}.
A key approach to improve \gls{oee} is to enable runtime control, managing the performance of the manufacturing system in real-time, including continuous monitoring and decision making~\cite{iannone2013managing}.

Decentralized methods allow for greater flexibility and adaptability in complex manufacturing environments and quickly respond to changes on the plant floor \cite{kovalenko2019model}. One such decentralized strategy is the use of \gls{mas} to control various types of complex systems. \gls{mas} has been applied to various aspects of manufacturing systems, such as product design, production planning, and control~\cite{lee2008multi}. These systems perform well in dynamic environments with increased uncertainty and complexity.

Despite the advantages of improved flexibility and adaptability, coordinating the activities of agents requires extensive communication and computational efforts during runtime with a shared understanding of the environment. To overcome these challenges, \gls{swt} have been proposed to support more efficient communication and knowledge sharing \cite{ocker2019framework}. Ontologies, enabled by \gls{swt}, play a crucial role in facilitating communication and interoperability in \glspl{mas}, particularly in the manufacturing domain~\cite{obitko2002ontologies}. \gls{swt} can provide improved communication and cooperation among agents by providing a common language and structure. 

While online communication load and computational effort in \gls{mas} have been reduced by using \gls{swt} \cite{ocker2019framework}, there are still challenges when ensuring effective runtime control of the \gls{mas} in manufacturing. 
One challenge lies in the ability of the \gls{mas} to adapt to changes in the production system and its environment, capturing all the necessary data for the agents to accurately measure the \gls{oee} of each machine.~\textit{(C1)}.
Another challenge involves the need for technical expertise for both customers and engineers to interact with agents in the \gls{mas}, as it requires a deep understanding of the system's architecture and communication protocols ~\textit{(C2)}.
There is also a challenge of continuously updating agents relevant to previous and current states of the resources/parts/processes in the system~\textit{(C3)}.

The main contribution of this work is the design of the ontology, i.e. \gls{kb}, for \gls{mas} in manufacturing to enable agents to access and update the \gls{kb} during runtime and provide continuous decision support based on a global context. The contributions of this paper over the previous work include: 
(1) an extendable \gls{kb} that effectively accumulates the \gls{mas}[’s] state and history of production to accurately measure the \gls{oee},
(2) an ontology-based interface, enabling customers and engineers to facilitate communication with the \gls{mas} during runtime, and
(3) the implementation of a mechanism where agents within the manufacturing system continuously update the \gls{kb} with their current state and the latest history of their actions.

The rest of the manuscript is organized as follows.
\Cref{sec:background} provides background regarding ontologies and \glspl{mas} in manufacturing.
\Cref{sec:framework} describes an ontology-based multi-agent framework to improve runtime control of manufacturing systems.
\Cref{sec:implementation} explains the implementation of the proposed framework
\Cref{sec:casestudy} presents a case study to demonstrate the proposed framework.
\Cref{sec:summary} summarizes the paper and discusses future work.

%% file: 20_stateoftheart.tex

Ontologies and multi-agent systems have been implemented in manufacturing systems to enable more flexible, adaptable, and efficient production systems.

\subsection{Ontologies in Manufacturing}
\label{subsec:sota-onto}

An ontology is understood as an ``explicit specification of a conceptualization'' \cite{gruber1993translation}. As such, ontologies can be used to represent knowledge about production systems. The ontology consists of the \gls{tbox}, the schema, and the \gls{abox}, which includes the instance level. Ontologies can be utilized to efficiently locate, categorize, represent, and reuse knowledge that already exists inside various information resources \cite{hadzic2009ontology}.

MASON (MAnufacturing’s Semantics ONtology) is one of the first examples of ontology in the manufacturing domain~\cite{lemaignan2006mason} to formalize and share data using the \gls{owl}. Ontologies have also been applied in specific applications within the manufacturing field. For example, Dinar et al. \cite{dinar2017design} proposed an ontology for Design for Additive Manufacturing (DFAM) to model 3D printed components. In the steel manufacturing industry, Common Reference Ontology for Steelmaking (CROS) was proposed as a solution to the problem of semantic interoperability \cite{cao2022core}.
One study presented an OWL-based manufacturing resource capability ontology (MaRCO) \cite{jarvenpaa2019development}, which aims to describe the capabilities of manufacturing resources in a common formal resource model. The project AutomationML ontology (AMLO) interlinks and integrates heterogeneous data in industrial systems design by providing a semantic representation of the AutomationML standard \cite{kovalenko2018automationml}. While these studies have developed ontologies that provide information about the combined capabilities of resources and integrate heterogeneous data, they do not focus on applications during runtime.

\subsection{Multi-Agent Systems in Manufacturing}
\label{subsec:sota-mas}

A variety of multi-agent architectures have been developed to achieve system-level control of manufacturing systems~\cite{Leitao2009b, leitao2013past, Yoon2008}. These multi-agent architectures use several software agents to make high-level decisions for different manufacturing system components \cite{kovalenko2019model, bi2021dynamic}. The high-level decisions made by the agents influence the overall performance of the manufacturing system \cite{barbosa2015dynamic}. Therefore, the design of these software agents is crucial in understanding and enhancing the performance of the manufacturing system.

Recently proposed \gls{mas} architectures for manufacturing systems contain instances of \glspl{pa} and \glspl{ra}~\cite{kovalenko2019model, Farid2015a, bi2023dynamic}. A \gls{pa} is an agent which makes decisions for a specific component in the production system. An \gls{ra} is a high-level controller for a resource (e.g., robot, machine) on the shop floor. An example proposed in~\cite{kovalenko2019model} presents a \gls{mas} architecture revolving around \glspl{pa} that have the ability to understand their surroundings, plan, and activities, and request necessary actions from the \glspl{ra}. Similarly, Bi et al. ~\cite{bi2023dynamic} developed a model-based \gls{ra} architecture that incorporates risk assessment and improves throughput in dynamic manufacturing environments. While these architectures provide data structures and requirements for coordinating the behavior of agents, their aim is not on utilizing the historical knowledge for improving the runtime control of the manufacturing system.

\subsection{Ontologies and Multi-Agent Systems in Manufacturing}
\label{subsec:sota-onto-mas}

The introduction of ontologies in \glspl{mas} allows agents to make decisions based on their shared understanding of the domain. Formalizing knowledge is key to enabling agents to dynamically adapt to system changes \cite{leitao2013past}. Thus, it has been suggested that the use of \glspl{swt} would improve the \gls{mas} of manufacturing systems \cite{obitko2002ontologies}.

\gls{adacor}~\cite{borgo2004role} is one of the early efforts for using an ontology and \gls{mas} in manufacturing systems. This project aims to develop an ontology for distributed manufacturing systems that include components and procedures to assist in scheduling and monitoring. 
Vrba et al. \cite{vrba2011a} suggested using a central ontology in the form of an \gls{rdf} database. A production plan agent provides each \gls{pa} with a possible production plan and then single agents schedule a plan by collaborating with available resources. 
An ontology framework for automatic initialization of multi-agent production systems using \gls{swt} was proposed in prior work \cite{ocker2019framework}. The ontology is effectively queried to produce an automaton for initializing the individual agents. As a result, the online communication load and the computational effort are reduced for \glspl{pa} and \glspl{ra}. However, in the framework proposed runtime information from the agents is not fed back to the ontology, which would allow for adapting and evolving over time as the environment changes.

In summary, several architectures have been proposed for using \gls{swt} and \gls{mas} for controlling manufacturing systems. However, none of these works aim to utilize an ontology for \gls{mas} in manufacturing to improve runtime control by feeding information back into the ontology. To the best of our knowledge, none of the existing works addresses the combination of utilizing an extendable \glspl{kb} to capture comprehensive historical information (\textit{C1}), providing an interface for customers and engineers for agent communication during runtime via the \gls{kb} (\textit{C2}), and updating the \gls{mas} during runtime (\textit{C3}) to provide decision support for the agents in a global context.

%% file: 30_framework.tex
In this section, we introduce an ontology-enhanced multi-agent framework for manufacturing that contains an extendable \gls{kb}, an interface for engineers and customers, and runtime control through communication and coordination between agents and the ontology.

\subsection{Design of the Framework}
\label{subsec:designOntology}

\begin{figure*}[t]
    \centering
    \captionsetup{belowskip=-7pt} 
    \includegraphics[width=.98\textwidth]{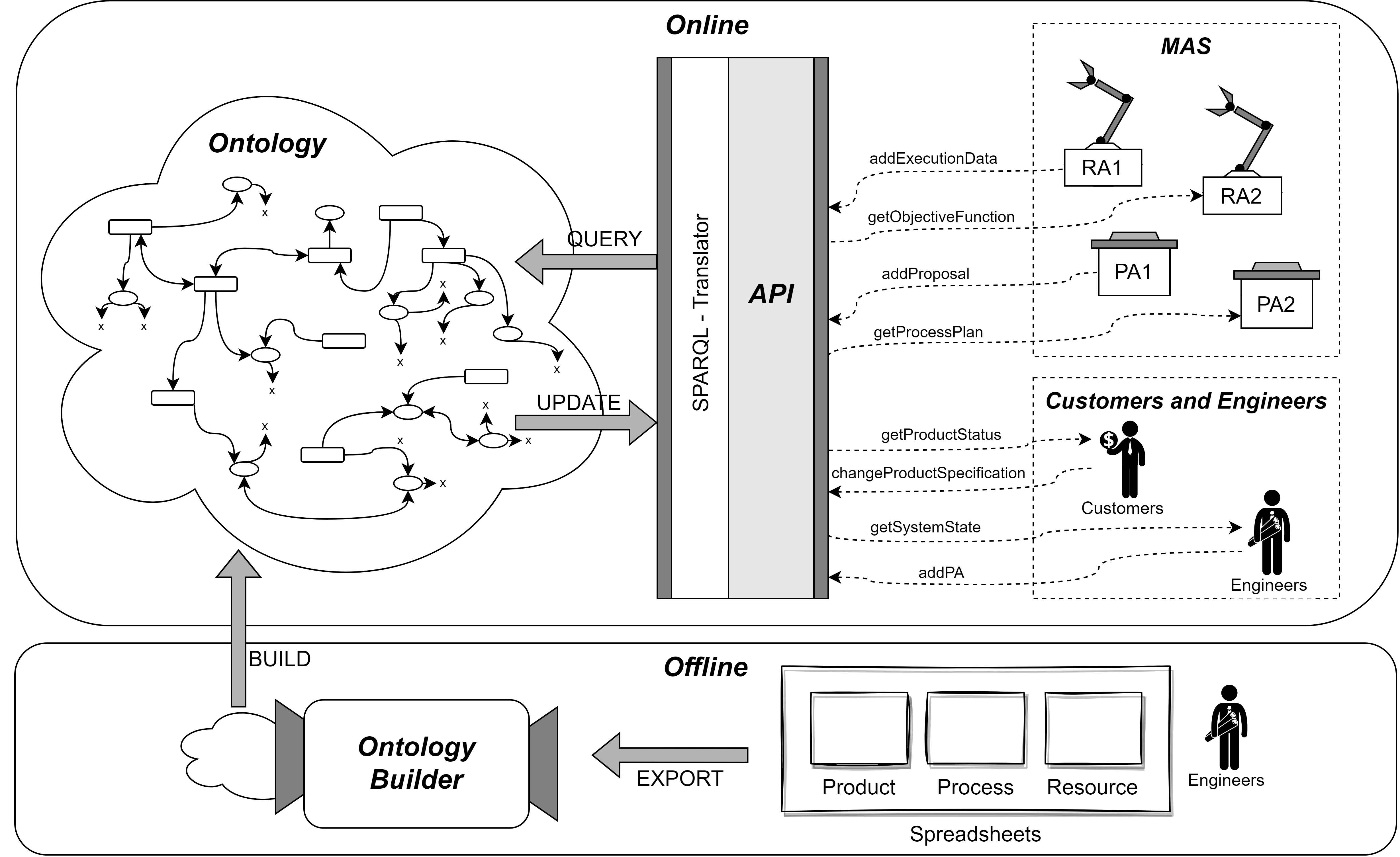}
    \caption{Graphical Abstract of the Framework}
    \label{fig:graphAbstract}
\end{figure*}

The novel architecture presented in \Cref{fig:graphAbstract} provides a general overview of an ontology-based multi-agent framework.
An ontology is designed based on the architecture of a \gls{mas} that includes both \glspl{pa} and \glspl{ra} using the \gls{bdi} paradigm. This ontology stores the knowledge about the production process, the resources, and the interactions. 
Section \ref{subsec:communication} offers a detailed discussion on how the knowledge is stored within the ontology. Engineers input the initial knowledge of a manufacturing system in the spreadsheets and then export it into the ontology using the \textit{Ontology Builder} during the ``offline'' phase, depicted in \Cref{fig:graphAbstract}.
Using the constructed ontology, agents in \gls{mas} communicate with the ontology through the \gls{api} during runtime, i.e. the ``online'' phase. The agents are indicated as \glspl{pa} and \glspl{ra} in the box of \gls{mas} in \Cref{fig:graphAbstract}. Both customers and engineers also have the ability to retrieve and update the \gls{kb} using the \gls{api} as shown in \Cref{fig:graphAbstract}. 
Some of the essential functions provided by the \gls{api} are introduced in Section \ref{subsec:runtimeControl}.

The \gls{bdi} architecture includes \textit{beliefs}, \textit{desires}, and \textit{intentions}, where \textit{beliefs} refer to the manufacturing environment, \textit{desires} define goals, and \textit{intentions} are plans derived from beliefs to fulfill desires \cite{kovalenko2019model}. Therefore, an ontology based on the \gls{bdi} architecture can provide a structured and organized way to represent the knowledge required for both \glspl{pa} and \glspl{ra} to operate effectively in the \gls{mas}.

The framework introduces three essential capabilities, offering main benefits over existing approaches.
First, the framework utilizes an ontology as a centralized \gls{kb} to accumulate the production history of the \gls{mas}. This knowledge provides agents with holistic environment models, offering a comprehensive understanding of the system's current state and history (\textit{C1}). The agents are able to make their respective decentralized decisions in a global context.
The framework also allows engineers and customers to easily communicate with the agents in the \gls{mas} through the use of ontology (\textit{C2}). They can retrieve and update the manufacturing environment without needing knowledge of the system architecture, thus providing flexibility and scalability.
Additionally, the framework provides capabilities for communication and coordination between agents and the ontology during the runtime of the manufacturing process (\textit{C3}). The agents can directly query the ontology for adjustments in response to global changes in resources, product features, or other factors.

 \subsection{The Basic Ontology}
\label{subsec:basicOntology}

\begin{figure}[t]
    \captionsetup{belowskip=-15pt} 
    \centering
    \includegraphics[width=.48\textwidth]{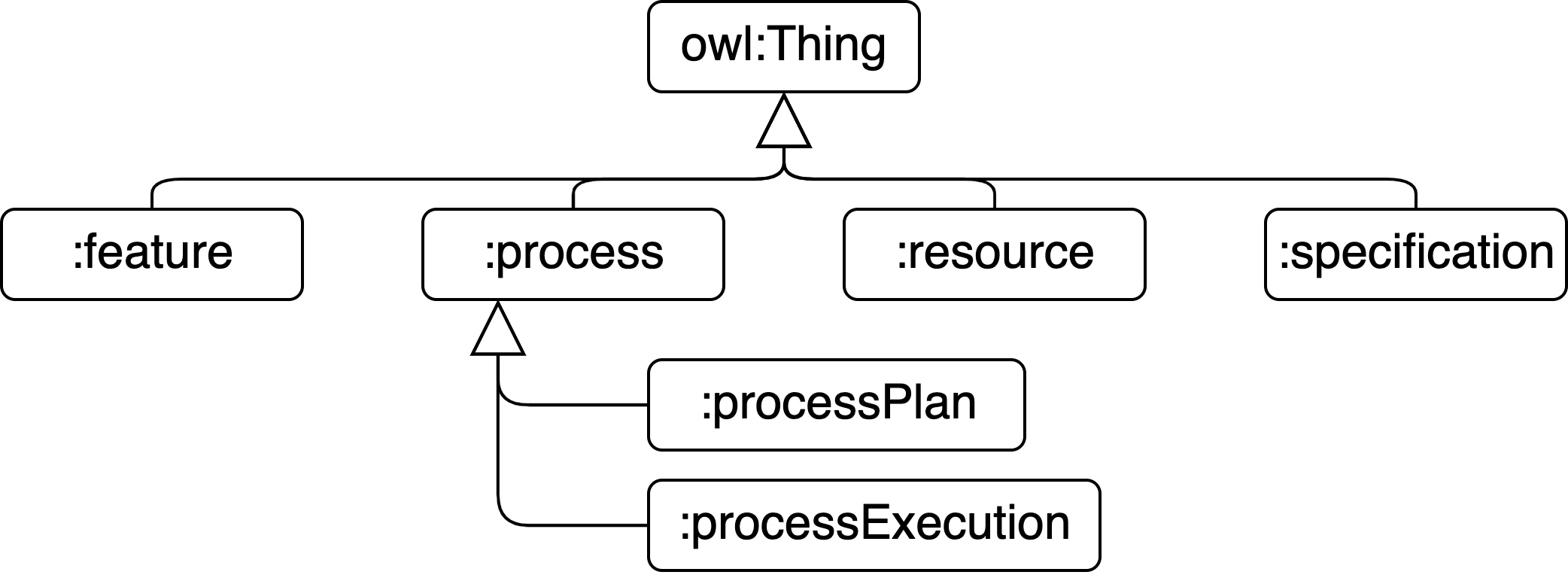}
    \caption{Overview of the Ontology}
    \label{fig:baseTbox}
\end{figure}

The core concepts in the manufacturing system that are defined in the ontology include these classes: \textit{feature}, \textit{process}, \textit{resource}, and \textit{specification}. This is based on the \gls{ppr} model, in which \textit{resources} execute \textit{processes} on \textit{products} to realize \textit{features} \cite{Nielsen2003}.

The \textit{specification} class refers to a detailed description of the desired product, including the specific features, requirements, and the expected deadline, as specified by the customer.
The \textit{process} class contains information on the production processes. These include both the \textit{processPlans}, which is a set of operations and procedures that transform inputs into finished products including the use of resources, as well as \textit{processExecutions}, which represent real-world execution of the respective process plans.
The \textit{resource} class encompasses physical assets or capabilities used in the production process such as robots and buffers.
The \textit{feature} refers to the various characteristics of a product that are defined in order and used to describe the desired product.
The initial TBox of the ontology that includes this information is shown in \Cref{fig:baseTbox}.

\begin{figure}[t]
    \centering
    \includegraphics[width=.48\textwidth]{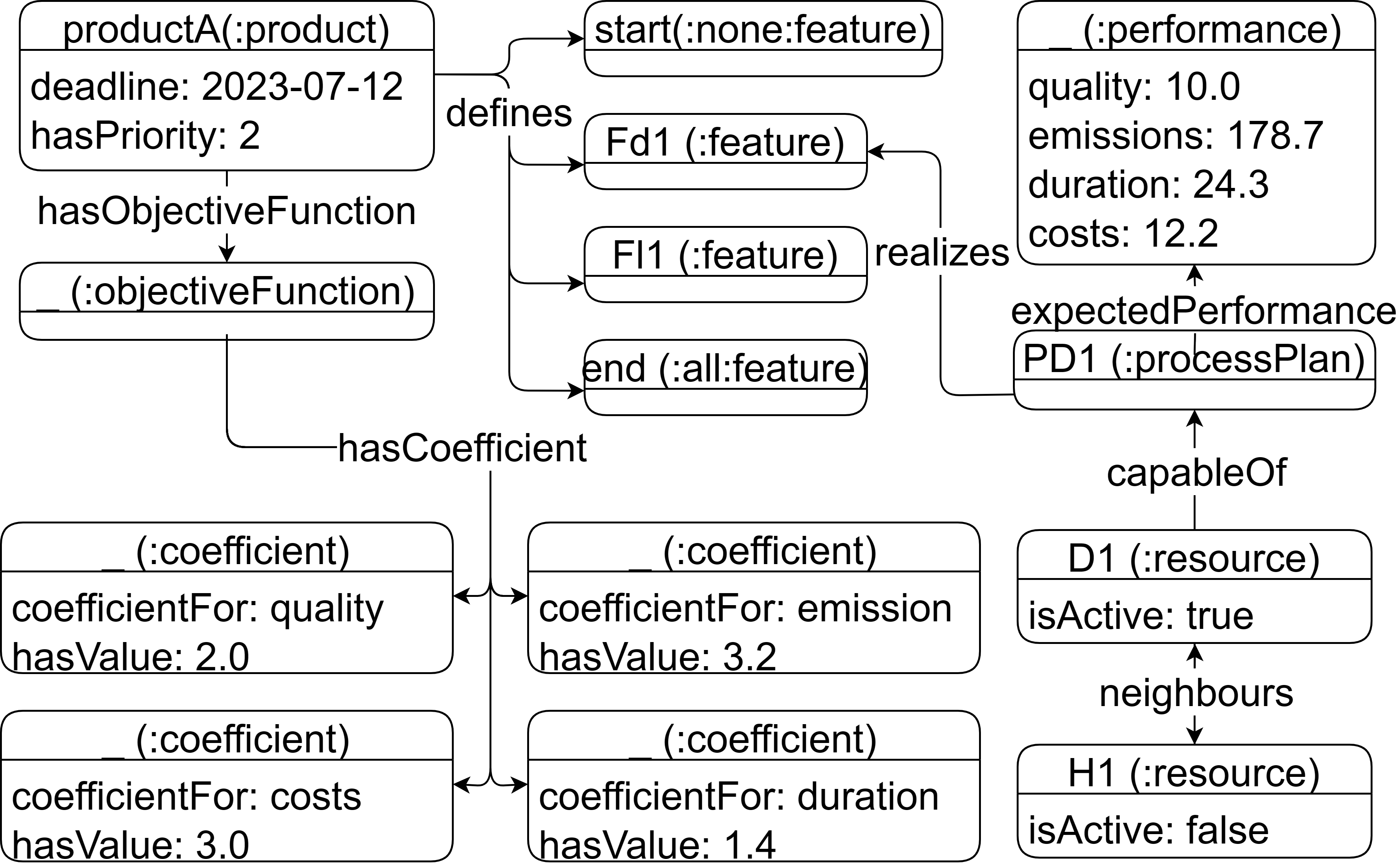}
    \caption{Relationship among Products, Processes, Features, and Resources}
    \label{fig:relationsAbox}
\end{figure}

To ensure that the \gls{kb} accurately represents not only the static \gls{mas}, but also the dynamically changing environment, further classes, and properties have to be defined. To align with the \gls{bdi} architecture and capture the \gls{pa} objectives, we propose an objective function that considers a set of use-case specific metrics which is shown in \Cref{fig:relationsAbox}. This is represented in the ontology by the \gls{pa} specific class \textit{objectiveFunction}. The link between the objective function and the \gls{pa} is created by the \textit{hasObjectiveFunction} property. The \textit{objectiveFunction} then has several \textit{coefficients} linked to it by the \textit{hasCoefficient} property. The value of the coefficient is given by the data property \textit{hasValue} and the type of metric by the \textit{coefficientFor} datatype property. Their respective performance also needs to be evaluated to allow for an evaluation of different process plans. This is done by introducing a \textit{performance} class and the \textit{hasPerformance} property. The \textit{hasPerformance} property has two subclasses, \textit{realPerformance} and \textit{expectedPerformance} to further enable transparency when it comes to historic decision-making.
The last metric for each product is the \textit{deadline}. By utilizing the previously mentioned classes, we can accurately represent both the beliefs (via environment models and expected performances) and desires (via objective function and deadline) of the \gls{bdi} architecture.

\subsection{Modeling the System during Runtime}
\label{subsec:runtimeControl}

\begin{figure}[t]
    \captionsetup{belowskip=-15pt} 
    \centering
    \includegraphics[width=.48\textwidth]{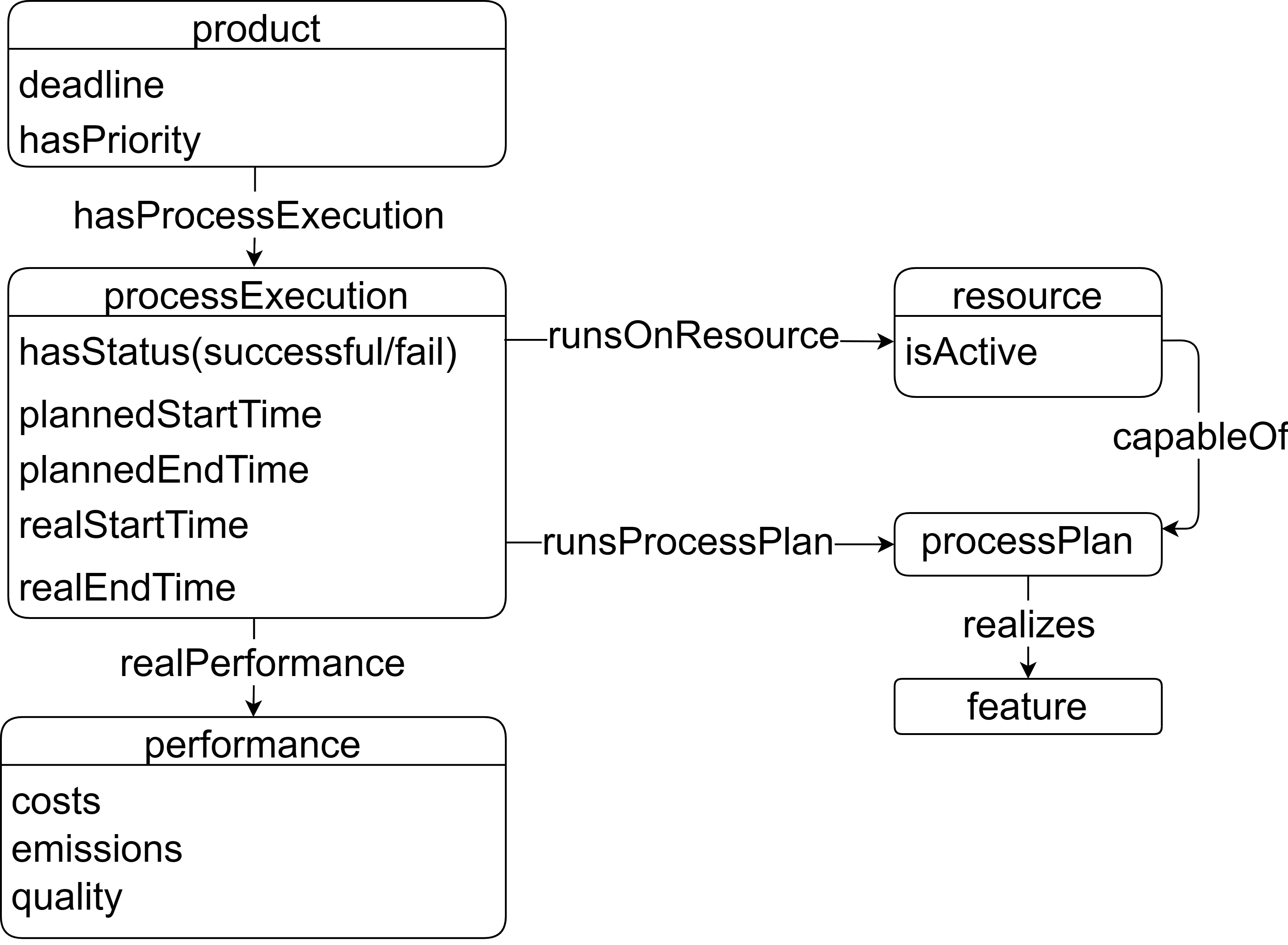}
    \caption{Example of Runtime Information Ontology}
    \label{fig:processExecution}
\end{figure}

To support runtime control of \gls{mas} utilizing the historical data, runtime information is added to the \gls{kb}. We developed a concept called \textit{processExecution}, establishing its relationship with product, resource, and performance, to provide agents information during runtime. A \textit{processExecution} represents a physical event, in which a \textit{processPlan} gets executed.

An example of runtime data is shown in \Cref{fig:processExecution}. A \gls{pa} adds each process execution to the ontology and displays its intention in a globally accessible format. To connect each process execution to different parts of the base ontology, a new set of object properties is introduced. The \textit{hasProcessExecution} property associates a product with execution, while the \textit{runsOnResource} property indicates which resource is utilized. The \textit{runsProcessPlan} property links the process plan to the execution.

The \textit{hasStatus} property captures the different stages of the execution by assigning a literal value. Initially, the \gls{pa} proposes the execution in the ontology and the status changes to \textit{planned}. When the execution starts, the status changes to \textit{running} and then to \textit{successful} if completed successfully. If unexpected events occur and the execution does not finish as planned, the status changes to \textit{errored} with a \textit{hasErrorMessage} property. Other important datatype properties for the execution include \textit{plannedStartTime} and \textit{plannedEndTime}, as well as \textit{realStartTime} and \textit{realEndTime}. The actual performance of the execution may differ from the expected performance of the process, which is captured by the \textit{realPerformance} property. By utilizing this model, we can incorporate past, present, and future events into an ontology, which further contributes to the belief component of the \gls{bdi} architecture. As a result, these events form the basis of intentions for the various agents within the \gls{mas}. The process execution model facilitates the integration and retrieval of a wide range of knowledge and capabilities during runtime. The following are some of the essential interactions implemented:

\begin{itemize}
  \item \textbf{addPlannedExecutionData}: This query adds data on the product that will be processed, planned start and end times, and the resources that will be utilized.
  \item \textbf{updateExecutionData}: This query updates execution data such as status, real start times, and end times.
  \item \textbf{getProductStatus}: This query is used to retrieve product-related information such as the status, duration, and deadline.
  \item \textbf{getResourceHistory}: This query allows retrieval of the historical data of the resources utilized in the manufacturing processes.
  \item \textbf{changeResourcePerformance}: This query allows modification of the performance such as energy costs, emissions, and quality.
\end{itemize}

By utilizing queries, various interactions can be implemented to retrieve and modify information such as resource history and performance. The agents can use this information to improve runtime control and \gls{oee} of machines in the system.

%% file: 40_implementation.tex
\subsection{Dynamically Building the Knowledge Base}
\label{subsec:buildingKB}

The approach for dynamically building the \gls{kb} is based on prior work \cite{ocker2019framework}. The initial ontology is created using spreadsheets to reduce dependence on proprietary tools. The spreadsheets are converted into CSV files using a macro. Finally, the transformation of these files into a formalized ontology using \gls{owl} is implemented in Python using the package Owlready2 \cite{owlready}. By leveraging this package, the ontology creation process was made more accessible. 

\subsection{Agent-Ontology Communication during Runtime}
\label{subsec:communication}

\begin{figure}[t]
    \captionsetup{belowskip=-15pt} 
    \centering
    \includegraphics[width=.48\textwidth]{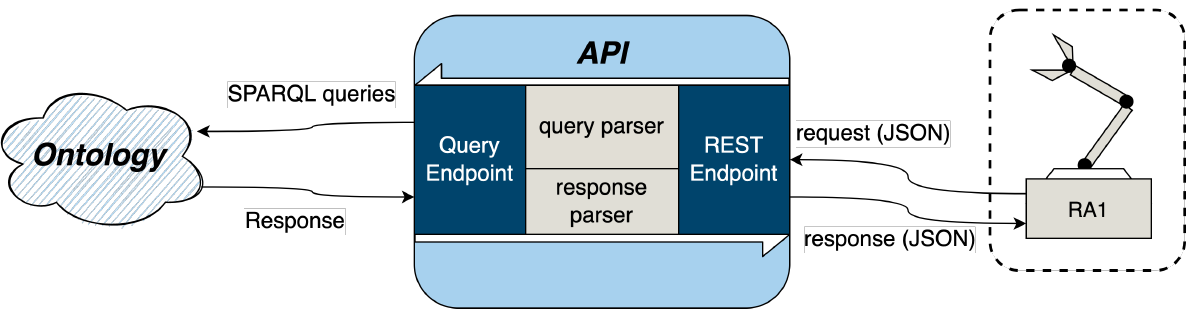}
    \caption{Communication between Agents and Ontology}
    \label{fig:agentCommunication}
\end{figure}

We developed an \gls{api} to facilitate communication between the \gls{mas} and the ontology. This \gls{api} allows interacting with the ontology, querying and updating the ontology using a \gls{http} requests. The \glspl{pa} and \glspl{ra} from the \gls{mas} request specific actions through this \gls{api}, which then utilizes a \gls{rest}-interface to pass the corresponding query to the ontology. 
When a \gls{pa} or \gls{ra} from the \gls{mas} retrieve or update information from an ontology, the agents send a request to the \gls{rest} \gls{api}. The \gls{api} then converts the request to \gls{sparql} queries, allowing agents to retrieve or manipulate data in the ontology.
We utilize the Stardog~\cite{Stardog} knowledge graph platform to run the \gls{sparql} queries outlined in \ref{subsec:runtimeControl}. For instance, if an agent requests information about a product's status, the \gls{api} will execute the appropriate \gls{sparql} query. The response is provided in \gls{json} to an agent. The format is optimized for the agents, meaning that it is structured in a way that makes it easy for the agents to use the data. Moreover, customers and engineers can also interact with the ontology to retrieve or update information during runtime. Note that in the context of this framework, there is no communication delay between the agents and ontology. A graphical representation of this communication between agents and ontology is provided in \Cref{fig:agentCommunication}.

\subsection{Querying the Knowledge Base}
\label{subsec:queryingKB}

\begin{figure}[t]
    \captionsetup{belowskip=-15pt} 
    \centering
    \includegraphics[width=.48\textwidth]{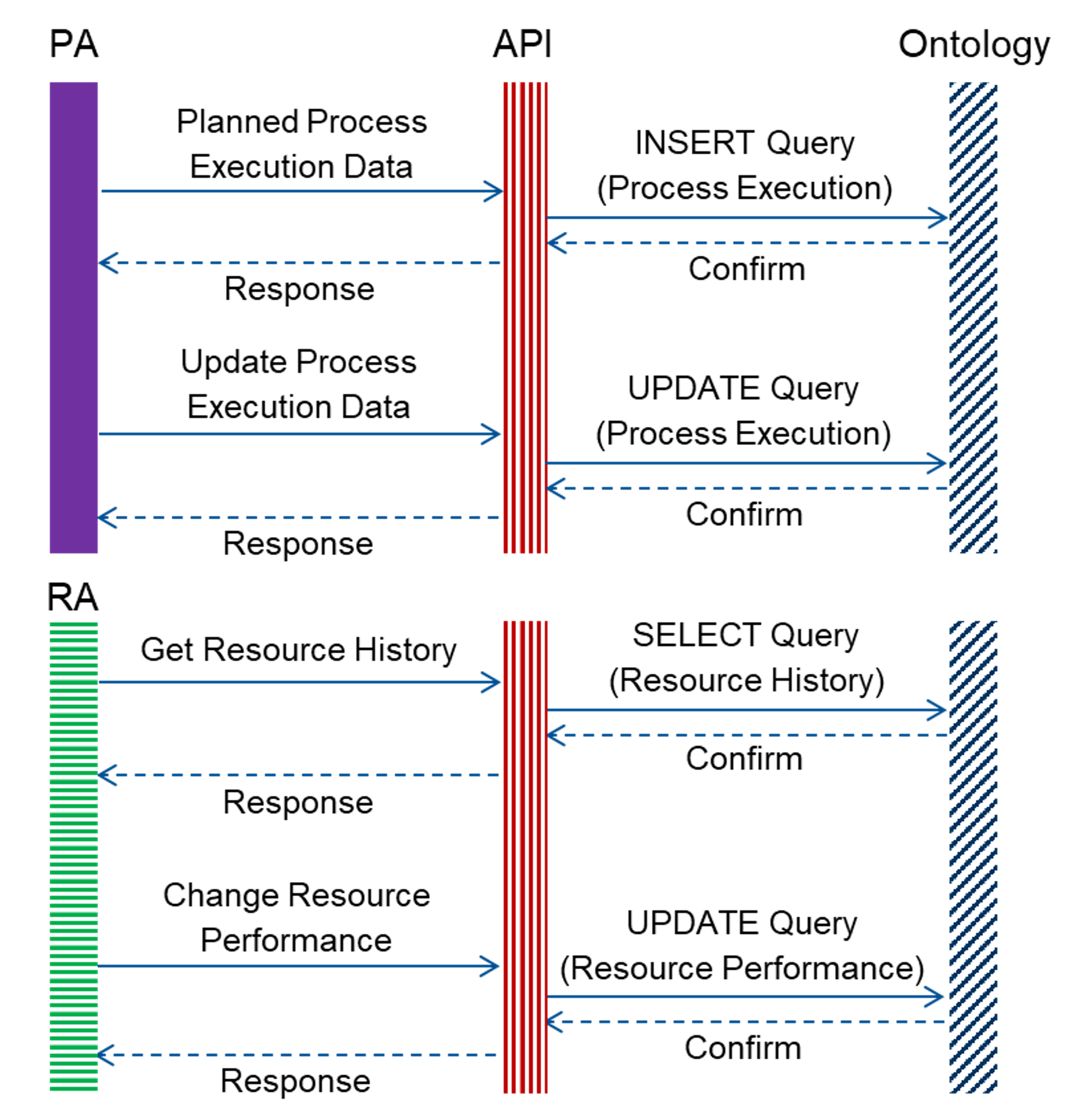}
    \caption{Querying the Knowledge Base}
    \label{fig:queryingKB}
\end{figure}

The ontology was designed to store information about processes, resources, relationships, and performance data. To register runtime information to an ontology based on this design, we utilize \gls{sparql} to interact with an ontology as described in Section \ref{subsec:communication}. The overall process of querying some of the knowledge and capabilities for a \gls{pa} and an \gls{ra} is shown in \Cref{fig:queryingKB}. 
The solid line represents the request flow from the agents to the ontology, while the dotted line represents the response flow from the ontology back to the agents.
An INSERT operation allows new data to be added, thus extending the \gls{kb}. For example, to register a new planned process execution, the data includes the properties of the execution such as its ID, part name, and planned start time and end time. 
For UPDATE operations, INSERT and DELETE operations can be combined.  When a process in the \gls{mas} is executed, we update the start time and status of a process.
The SELECT operation can retrieve information during runtime. For example, an \gls{ra} can use this operation to retrieve information about the resource's past activities. The \gls{sparql} query illustrated in \Cref{lst:sparql} retrieves information associated with successful process executions run on a particular resource. The query selects distinct values for the execution ID, emissions, costs, quality, and start and end time. 
The question mark (``?'') is used as a prefix to represent variables that will be assigned values during the execution of the query.
It accomplishes this by searching for a resource identified by the <resource> URI and filtering for executions with a status of "successful". The query then looks for performance data including emissions, costs, and quality. This information can be used by an \gls{ra} to make decisions about allocating resources to improve the manufacturing process.

\lstset
{
    basicstyle=\footnotesize,
    numbers=left,
    stepnumber=1,
    showstringspaces=false,
    tabsize=1,
    breaklines=true,
    breakatwhitespace=false,
    xleftmargin=8.0ex,
    frame=lines,
    framexleftmargin=8.0ex,
}
\lstdefinelanguage{sparql}{
morestring=[b][\color{blue}]\",
keywordstyle=\color{tumblue},
morekeywords={SELECT,CONSTRUCT,DESCRIBE,ASK,WHERE,FROM,NAMED,PREFIX,BASE,OPTIONAL,FILTER,GRAPH,LIMIT,OFFSET,SERVICE,UNION,EXISTS,NOT,BINDINGS,MINUS,a,DISTINCT,ORDER,GROUP,BY,COUNT,AS,UNION,INTERSECTION,HAVING,BIND},
sensitive=true
}

\noindent
\begin{minipage}{\columnwidth}
\begin{lstlisting}[language=sparql, caption={SPARQL Query for retrieving a resource's history.},label=lst:sparql]
SELECT DISTINCT ?execution ?emissions ?costs 
                ?quality ?startTime ?endTime
 WHERE
{
    ?execution :runsOnResource <resource>;
               :hasStatus "successful";
               :realPerformance [
                 :emissions ?emissions;
                 :costs ?costs;
                 :quality ?quality
               ];
               :realStartTime ?startTime;
               :realEndTime ?endTime.
} ORDER BY ?startTime
\end{lstlisting}
\end{minipage}

%% file: 50_casestudy.tex
In this case study, we examine the proposed framework and demonstrate how the \gls{mas} utilizes the ontology to improve runtime control.

\subsection{Case Study Setup and Workflow}
\label{subsec:setup}

\begin{figure}[t]
    \captionsetup{belowskip=-15pt} 
    \centering
    \includegraphics[width=.48\textwidth]{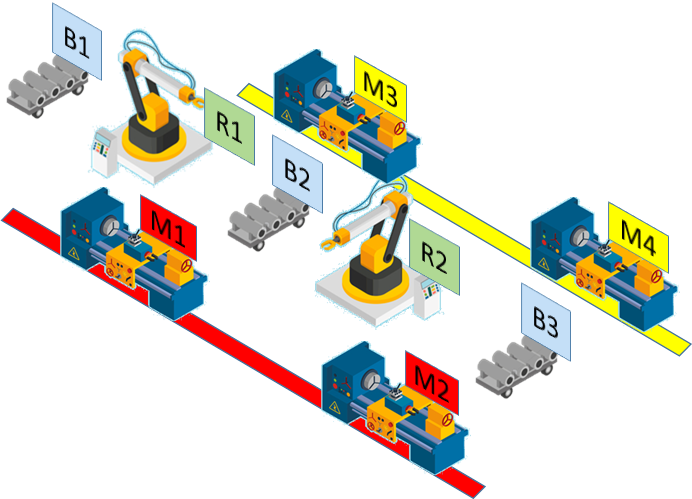}
    \caption{Layout of a Small Manufacturing System}
    \label{fig:plantLayout}
\end{figure}

To evaluate the feasibility, a case study is set up in a small manufacturing system environment, which is composed of four identical machines (M1, M2, M3, and M4), two robots (R1 and R2), and three buffers (B1, B2, and B3). The system has 6 \glspl{ra} representing R1, R2, M1, M2, M3, and M4. R1 transfers the part from B1 to B2. R2 transfers the part from B2 to a machine. When a manufacturing process P1 is completed, a part is transferred from a machine to B3 to exit the system. 
The performance of the machines is shown in \Cref{table_initial_setup}.

The objective of \gls{pa} and \gls{ra} is to enhance the runtime control of \gls{mas} in manufacturing by utilizing the knowledge in the ontology. Specifically, the goal of \glspl{pa} is to fulfill the product requirements based on the \textit{objectiveFunction}, which is described in Section \ref{subsec:basicOntology}.
The goal of \glspl{ra} is to increase their \gls{oee} while limiting energy utilization. In this example, the \glspl{ra} try to increase their OEE by ensuring that each resource has an uptime over 50\% while limiting the total energy cost to 450 kWh.
Each machine's performance is adjusted by controlling the duration and energy cost by the \glspl{ra}. In this case study, the assumption is made that 5\% of energy cost from the previous cost results in a one-minute decrease in duration. The layout of a small manufacturing system is depicted in \Cref{fig:plantLayout}.

A new manufacturing process is initiated once a new part enters the system beginning at B1. As a part enters the system, product data are initialized in the ontology as shown in \Cref{fig:relationsAbox}. This data also contains information such as planned start and end time, provided in \Cref{fig:processExecution}. When R1 \gls{ra} queries from the ontology that a process P1 is planned, R1 moves the part from B1 to B2. Once the part has been moved to B2, the \gls{pa} updates the data for P1 by specifying the machine to be run on. It should be noted that the ontology only provides each \gls{pa} with possible plans, and does not compute optimal plans. Therefore, the \gls{pa} identifies the machine to be used by evaluating the \textit{objectiveFunction}, as depicted in \Cref{fig:relationsAbox}. Then, the \gls{pa} selects a machine that allows parts to exit the system as fast as possible, taking into account the time required for previously running parts inside the machines. When R2 \gls{ra} detects from the ontology that a part is planned to be processed on a machine such as M2, R2 moves the part to M2. After the part is processed, the \gls{pa} updates the ontology by setting the \textit{hasStatus} to ``successful'' and \textit{realEndTime} to the current time using the UPDATE Query shown in \Cref{fig:queryingKB}. Finally, R2 picks up the part from M2 to B3 to exit the system.

\begin{table}[t]
\caption{Peformance of the Machines}
\label{table_initial_setup}
\begin{center}
\begin{tabular}{|c|c|c|c|c|}
\hline
Machines & M1 & M2 & M3 & M4 \\
\hline
Duration (minutes) & 20 & 18 & 15 & 17 \\
\hline
Energy Costs (kWh) & 100 & 110 & 120 & 115 \\
\hline
\end{tabular}
\end{center}
\vspace{-15pt} 
\end{table}

\subsection{Case Study Scenario}
\label{subsec:description}

In the case study, we created a scenario of how \glspl{ra} use the ontology to adjust the performance levels of each machine. The \glspl{ra} of each machine queries the ontology to obtain historical data about the status, energy costs, quality, and process time of each machine. Specifically, the \glspl{ra} query the ontology using the \gls{sparql} query shown in \Cref{lst:sparql} to retrieve the performance of each completed process for their respective machines. The query provides \glspl{ra} with the uptime and real performance over the expected performance data of each machine. 

Suppose that after an initial set of operations, the uptime of M1 is below the threshold of 50\% (e.g. M1: 48\%, M2: 60\% M3: 80\% M4: 70\%) while the performance efficiency was the best at M1 (e.g. M1: 95\%, M2: 93\% M3: 80\% M4: 85\%). To improve the uptime of M1, The M1 \gls{ra} increases the energy cost and reduces the duration, while the M3 \gls{ra} decreases the energy cost and increases the duration. For example, the \gls{ra} of M1 reduces the duration of M1 from 20 to 18 minutes and increases the energy cost by 10.25\% (from 100 to 110.25) while M3 increases the duration from 15 to 16 minutes and decreases the energy by 5\% (from 120 to 114) using the \gls{api} function called \textit{changeResourcePerformance}. With these adjustments, the uptime of M1 not only exceeds 50\%, leading to improvements in \gls{oee} but also ensures that the products are processed in the most high-performing machine available, thereby maximizing productivity.

In this case study, we showcase how a small manufacturing system exhibits nominal behavior by utilizing ontology for monitoring the system's status in a global context. The agents involved in the system continuously monitor the ontology during runtime to make informed decisions, leading to improved runtime control of the manufacturing process.

%% file: 60_summary.tex
In this paper, we presented an ontology-based approach for implementing \gls{mas} in manufacturing to enable agents and users to access and update the \gls{kb} during runtime. The framework proposed for implementing an ontology for \gls{mas} in manufacturing has several advantages over existing approaches. First, the extendable \gls{kb} accumulates the system's state and history of production data, enabling agents to make informed decisions based on this data to measure the \gls{oee} (\textit{C1}). Second, providing an \gls{api} enables customers and engineers to communicate with the \gls{mas} during runtime without requiring detailed technical knowledge of the system (\textit{C2}). Lastly, the agents are continuously updating the \gls{kb} with their current state and their latest actions, which lead to improvements of \gls{oee} (\textit{C3}).

Future research will include evaluating this ontology-based framework by incorporating additional criteria that focus on improving system efficiency and effectiveness to further validate its benefits over existing approaches. Also, we plan to explore the application of this ontology-based framework to a real-world testbed, aiming to achieve improved \gls{oee} through ontology utilization during runtime with \gls{mas}. This testbed will enable us to prove the effectiveness of the framework in a practical manufacturing environment.